# Slow Magnetic Relaxation of a 12-Metallacrown-4 Complex with a Manganese(III)-Copper(II) Heterometallic Ring Motif

*Alex J. Lewis[†], Elena Garlatti[§‡], Francesco Cugini[‡], Massimo Solzi[‡], Matthias Zeller[¥], Stefano Carretta*[§‡], and Curtis M. Zaleski[†]**

[†]Department of Chemistry and Biochemistry, Shippensburg University, Shippensburg, Pennsylvania 17257, United States

[‡]Dipartimento di Scienze Matematiche, Fisiche e Informatiche, Università di Parma, 1-43124 Parma, Italy

[§]Udr Parma, INSTM, 1-43124 Parma, Italy

[¥]Department of Chemistry, Purdue University, West Lafayette, Indiana 47907, United States

ABSTRACT

The heterobimetallic metallacrown (MC), $(TMA)_2\{Mn(OAc)_2[12-MC_{Mn^{III}Cu^{II}(N)shi}-4](CH_3OH)\}\cdot2.90CH_3OH$, **1**, where $TMA^+$ is tetramethylammonium, $^-OAc$ is acetate, and $shi^{3-}$ is salicylhydroximate, consists of a $Mn^{II}$ ion captured in the central cavity and alternating unambiguous and ordered manganese(III) and copper(II) sites about the MC ring, a first for the archetypal MC structure design. DC-magnetometry characterization and subsequent simulation with the Spin Hamiltonian $\mathcal{H} = -J_1(\boldsymbol{s}_1 + \boldsymbol{s}_3) \cdot \boldsymbol{s}_5 - J_2(\boldsymbol{s}_2 + \boldsymbol{s}_4) \cdot \boldsymbol{s}_5 - J_3 \sum_{i=1}^{4} \boldsymbol{s}_i \cdot \boldsymbol{s}_{i+1} + d\,(s_{z,1}^2 + s_{z,3}^2) + \mu_B \sum_{j=1}^{5} g_j \boldsymbol{s}_j \cdot \boldsymbol{B}$, indicate an S = 5/2 ground state and a sizeable axial zero-field splitting on $Mn^{III}$. AC-susceptibility measurements reveal that **1** displays slow magnetization relaxation akin to single-molecule magnet (SMM) behavior.

TEXT

Metallamacrocyclic compounds such as wheels[1], rings[2], helicates[3], cryptates[4], calixareneses[5], coronates[6], and metallacrowns[7], provide a great deal of synthetic flexibility for chemists. Indeed, these molecules can be varied by substituting one metal[8] or ligand[9] choice for another without changing the overall molecular structure. Thus, the physical properties of a system can be fine-tuned by component choice. The characteristics and properties of these compounds impact a large breath of areas, including metalloenzyme active sites models, luminescence, molecular magnetism, and sensors.[10] In particular, metallacrowns (MCs) have proven to be a versatile class of coordination complexes, from structural and functional analogues of crown ethers to recent examples of bioimaging agents.[11,12] Furthermore, MCs tend to possess interesting magnetic properties, especially as SMMs.[12-14] Their wide range of applications is due to their ability to be formed in a predictable fashion based on metal and ligand choice and that the systems can incorporate different metal types in one molecule.

The archetypal MC motif consists of a cyclic ring with a M-N-O repeat unit and the oxygen atoms generate a central cavity, which binds a metal ion. Though the definition of a MC has expanded since the first reports to include molecules such as azametallacrowns, inverse MCs, collapsed MCs, and other larger structures that do not strictly adhere to the M-N-O repeat unit.[12,13]

While the first MCs tended to be homometallic, chemists quickly devised ways to synthesize heterobimetallic systems and since 2014 heterotrimetallic MCs have become more common.[15] In most heterometallic MC systems, the ring metals are of one type e.g. a transition metal, and the metal ion(s) captured in the central cavity is of a different variety (an alkali and/or lanthanide ion). What is lacking in the field is the use of two or more different $3d$ transition metal ions in an archetypal MC complex. Only one example qualifies as an archetypal MC with more than one $3d$ transition metal: a 12-MC-4 complex with ring $Fe^{III}$ ions and a central $Cu^{II}$ ion.[16] This molecule is though typical of other heterobimetallic systems with segregated sites for the two different metal species.

Herein we report the first archetypal MC, $(TMA)_2\{Mn(OAc)_2[12\text{-}MC_{Mn^{III}Cu^{II}(N)shi}\text{-}4](CH_3OH)\}\cdot 2.90CH_3OH$, **1**, to contain two different metal ions in the MC ring with manganese(III) and copper(II) ions in unambiguous coordination sites. This is significant as two different metal ions are often disordered about the coordination sites in other heterometallic macrocycles, such as metallic wheels and rings.[8,17] By careful choice of ligands and synthetic schemes ordered heterometallic antiferromagnetic rings can be obtained[9,18]; however, the majority of instances have a disordered metal sites. For the $3d$ heterobimetallic $12\text{-}MC_{Mn^{III}Cu^{II}}\text{-}4$ compound **1**, the manganese(III) and copper(II) ions alternate in a defined, ordered fashion about the ring. Additionally, DC-magnetometry characterization indicates a mixed antiferromagnetic-ferromagnetic exchange between the metals, and AC-magnetic susceptibility measurements suggest SMM-like behavior.

Complex **1** can be prepared by combining salicylhydroxamic acid and tetramethylammonium acetate in methanol followed by the addition of copper(II) acetate and manganese(II) acetate (synthetic details in SI). Complex **1** possesses the traditional metallacrown connectivity with a M-N-O repeat unit about the MC ring, and the $Mn^{III}$-N-O-$Cu^{II}$-N-O pattern recurs twice to generate the MC (Figures 1 and S1). The metallacrown captures a $Mn^{II}$ ion in the central cavity akin to the original 12-MC-4: $Mn(OAc)_2[12-MC_{Mn^{III}(N)shi}-4](DMF)_6\cdot 2DMF$, **2**, which contains a central $Mn^{II}$ and only ring $Mn^{III}$ ions.[11,19] If the expanded definitions of MCs are considered, several heterometallic ring MCs have been reported including collapsed MCs[20], inverse MCs[21], and large MC-like complexes[22]. However, these structures either lack a central cavity, bind a nonmetal atom in the central cavity, or lack a continuous M-N-O repeat unit, respectively. None of these systems are the archetypal MC with a M-N-O connectivity throughout the molecule and a central metal ion captured in the MC cavity.[12,13] Thus, **1** represents the only heterometallic ring archetypal MC reported to date. Moreover, the two different ions are ordered about the MC ring unlike many heterometallic ring compounds.[8,17]

The metal oxidation states are based on average bond lengths, bond valence sum (BVS) values[23], and overall molecular charge balance considerations (Tables S2 and S3). Four triply deprotonated salicylhydroximate ligands and two acetate anions provide a 14- charge, which is counterbalanced by two ring $Mn^{III}$ ions, two ring $Cu^{II}$ ions, one central $Mn^{II}$ ion, and two lattice tetramethylammonium cations (14+ charge). The overall structure of the MC is a square molecule that is slightly domed away from the central $Mn^{II}$ ion. The $Mn^{II}$ (Mn3) is six-coordinate with a coordination sphere comprised of four oxime oxygen atoms from four shi$^{3-}$ ligands, which yield the MC cavity, and two carboxylate oxygen atoms from two acetates. The acetates serve to link the central $Mn^{II}$ to each ring $Mn^{III}$ ion. A SHAPE analysis (*SHAPE 2.1*; Table S4) of the $Mn^{II}$ geometry yields the lowest continuous shape measure (CShM) for an octahedron; however, the geometry is severely distorted as the CShM value (3.491) is greater than the upper threshold value (3.0) that is considered as an adequate description of the geometry.[24]

The ordered nature of the $Cu^{II}$ and $Mn^{III}$ ions about the MC ring is likely due to the preferred coordination environment of each ion. Furthermore, BVS values and single-crystal X-ray analysis (details in SI) confirm the unambiguous assignment and ordered nature of the ring metal sites. Both $Cu^{II}$ ions are four-coordinate and possess a square planar geometry (Table S4). The coordination environments of Cu1 and Cu2 are identical and consist of trans-chelate rings. A six-membered chelate ring is formed by the oxime nitrogen and phenolate oxygen atoms of one $shi^{3-}$ and the opposite five-membered chelate ring is formed by the oxime and carboxylate oxygen atoms of another $shi^{3-}$. The two ring $Mn^{III}$ ions possess different coordination environments. Mn1 is six-coordinate and has an octahedral geometry with an elongated Jahn-Teller (JT) axis typical for high-spin $d^4$ ions (Table S4). The equatorial plane consists of trans five- and six-membered chelate rings, same as the $Cu^{II}$ ions. The axial ligands consist of a carboxylate oxygen atom of a bridging acetate and an oxygen atom from a methanol molecule. Mn2 is five-coordinate and possesses a square pyramidal geometry (Table S4) and the tau ($\tau$) value of 0.103 supports the assignment.[25] The basal ligands consist of trans five- and six-membered chelate rings, and the apical ligand is a carboxylate oxygen atom of a bridging acetate.

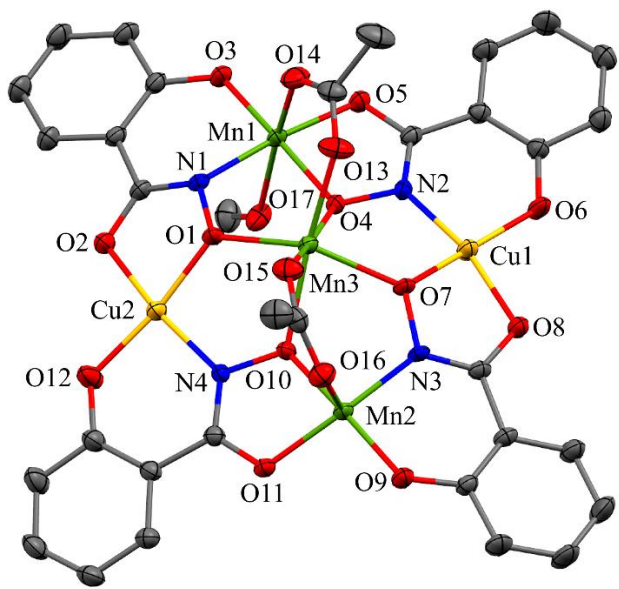

Figure 1. Single-crystal X-ray structure of $(TMA)_2\{Mn(OAc)_2[12\text{-}MC_{Mn^{III}Cu^{II}(N)shi}\text{-}4](CH_3OH)\}\cdot2.90CH_3OH$, **1**. The ellipsoid plot is at the 50% level. Hydrogen atoms, tetramethylammonium cations, and solvent molecules have been omitted for clarity. Color scheme: green, manganese; yellow, copper; red, oxygen; blue, nitrogen; gray, carbon.

Variable-temperature dc magnetic susceptibility measurements of **1** indicate the presence of sizeable antiferromagnetic (AF) interactions (Figure 2). The room temperature $\chi T$ value (9.1 emu K mol⁻¹) is less than that expected for the isolated ions (11.13 emu K mol⁻¹) and decreases with decreasing temperature. The low temperature $\chi T$ value (~3.2 emu K mol⁻¹ at 2 K) suggests a total-spin S = 5/2 ground state, split by sizeable zero-field splitting (ZFS) effects. The following Spin Hamiltonian can model the magnetic properties of **1**:

$$\mathcal{H} = -J_1(\boldsymbol{s}_1 + \boldsymbol{s}_3)\cdot\boldsymbol{s}_5 - J_2(\boldsymbol{s}_2 + \boldsymbol{s}_4)\cdot\boldsymbol{s}_5 - J_3\sum_{i=1}^{4}\boldsymbol{s}_i\cdot\boldsymbol{s}_{i+1} + d\,(s_{z,1}^2 + s_{z,3}^2) + \mu_B\sum_{j=1}^{5}g_j\boldsymbol{s}_j\cdot\boldsymbol{B}, \qquad (1)$$

(with $s_1 = s_3 = 2$ for $Mn^{III}$, $s_2 = s_4 = 1/2$ for $Cu^{II}$, $s_5 = 5/2$ for $Mn^{II}$ and $i + 1 = 1$ for $i = 4$). The first three terms of (1) represent the isotropic Heisenberg exchange interactions (inset Figure 2), the second term is the axial ZFS on the $Mn^{III}$ ions, and the last term accounts for the interaction with the applied magnetic field. The code PHI[26] was used to simultaneously fit susceptibility and magnetization data at T = 2 K (Figure 3), and two different exchange models, both with axial ZFS on the $Mn^{III}$ ions, were considered: one with two couplings constants (2J model, $J_1 = J_2$, $J_3$) and another with three couplings constants (3J model, with $J_1 \neq J_2$, $J_3$).

For the 2J model, the ring $Mn^{III}$-$Cu^{II}$ ions have a ferromagnetic (FM) interaction $J_3$ and they are AF-coupled with the central $Mn^{II}$. Indeed, $Mn^{III}$-$Cu^{II}$ compounds with oxime-based bridging ligands can display FM interactions[27], especially in presence of an elongated JT distortion for $Mn^{III}$ as in **1**.[28] The number of free parameters can be reduced by setting $|J_{1,2}| = |J_3|$ without affecting the agreement between the data and the simulation of the magnetic susceptibility (Figure 2, red circles) and the magnetization at 2 K (Figure 3, red circles) with $J_1 = J_2 = -12.5 \pm 1.6$ cm$^{-1}$, $J_3 = 12.5 \pm 1.6$ cm$^{-1}$ and $d = -6.5 \pm 1$ cm$^{-1}$. For the simulation, $g_{Mn^{II}}$ equaled 2, a typical value for S-ions, and $g_{Cu}$ and $g_{Mn^{III}}$ were allowed to vary around typical values[29] ($g_{Cu}$ = 2.2 and $g_{Mn^{III}}$ = 1.96). Energies levels of the exchange and the full Spin Hamiltonian (1) as a function of the magnetic field for the 2J model are found in Figure 4.

With an additional free parameter (3J model) it is possible to improve the agreement with experimental data which leads to two additional sets of best-fit parameters: (1) AF $J_3 = -13.2$ cm$^{-1}$, $J_1 = -7.7$ cm$^{-1}$, $J_2 = -27.1$ cm$^{-1}$ and $d = -6.4$ cm$^{-1}$ (Figures 2 and 3, green triangles) and (2) FM $J_3 = 2.1$ cm$^{-1}$, $J_1 = -6.2$ cm$^{-1}$, $J_2 = -48.6$ cm$^{-1}$ and $d = -7.5$ cm$^{-1}$ (Figures 2 and 3, blue diamonds). Energies of the total-spin multiplets calculated with the 3J models are reported in Figure S2.

In the model for **1**, only Mn1 and Mn2 have a sizeable single-ion ZFS, within the range of values expected for Mn$^{III}$ ions. To minimize the number of parameters, the same $d$ value was assumed for both, even though they have different coordination geometries. All considered models lead to an axial ZFS parameter $d$ of about -7 cm$^{-1}$. Additionally, no further improvement of the low-temperature susceptibility or magnetization simulations can be obtained by adding a rhombic term for the ZFS of the Mn$^{III}$ ions or by assuming a small (~1 cm$^{-1}$) ZFS on Mn$^{II}$.

Given the available data, the simplest and most reasonable model for **1** is the 2J model, with FM coupling between the ring Mn$^{III}$ and Cu$^{II}$ ions. The 2J model reduces the number of degrees of freedom without significantly affecting the agreement with the data, and the model is supported by previous results that indicate FM couplings between Mn$^{III}$ and Cu$^{II}$ ions mediated by N-O bridges.[27,28]

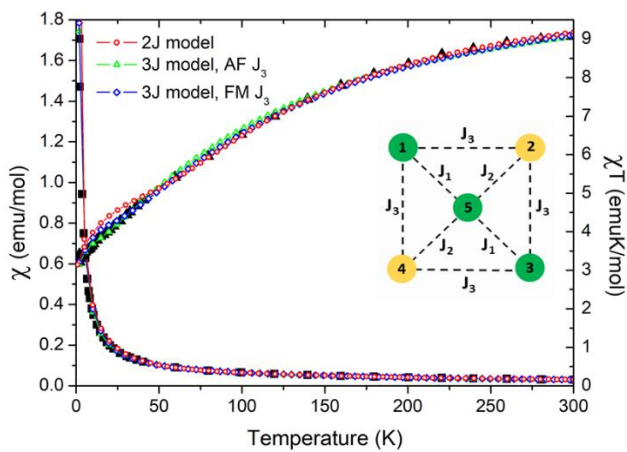

Figure 2. Temperature-dependence of the magnetic susceptibility ($\chi$; black squares) and $\chi T$ (black triangles) of **1** (B = 1000 Oe). Line/scatter represent best-fit calculations with the 2J model (red

circles) and the two 3J models (green triangles, blue diamonds). Inset: schematic representation of the exchange couplings.

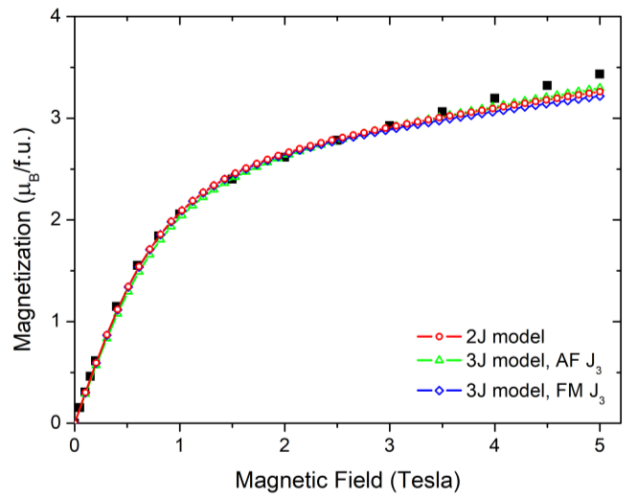

Figure 3. Field-dependence of the magnetization of **1** (squares) at T = 2 K. Line/scatter represent the best-fit to the experimental data with the 2J model (red circles) and the two 3J models (green triangles, blue diamonds).

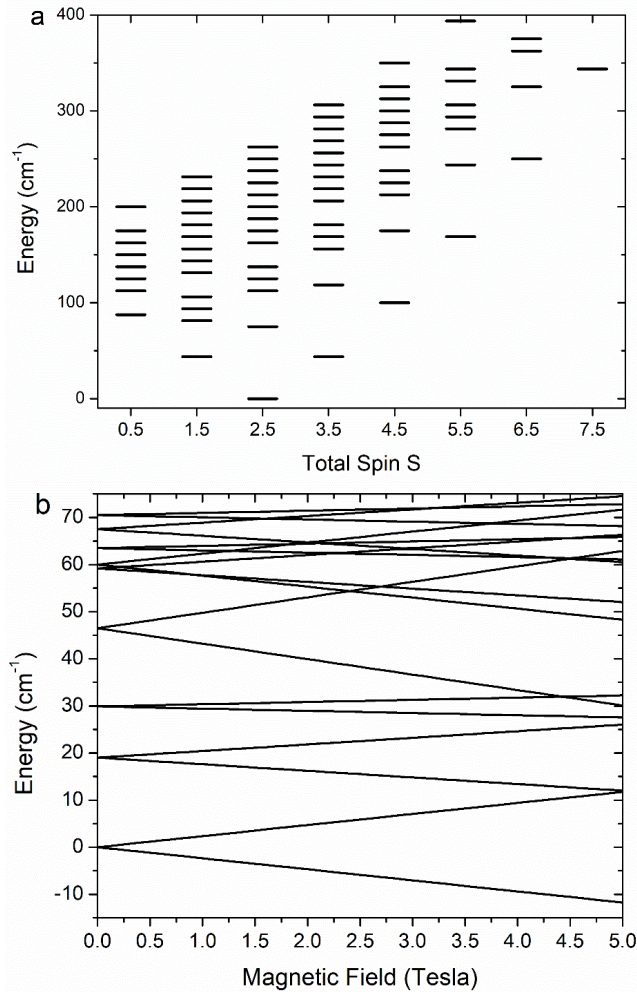

Figure 4. (a) Energy of the total-spin multiplets of **1** calculated with the exchange-only Spin Hamiltonian (S2) and the best-fit constants of the 2J model. The S = 5/2 GS energy is set to zero, with degenerate first-excited multiplets S = 3/2 and S = 7/2. (b) Magnetic field dependence of the lowest lying energy levels calculated with the full Spin Hamiltonian (1) and the 2J model. The magnetic field is applied along the *z* axis.

The dynamic magnetic behavior of **1** was probed by measuring the AC susceptibility at different frequencies (100 – 1000 Hz) with zero applied dc magnetic field (Figure S3). A frequency-dependent in-phase ($\chi'$) signal was observed below 3 K, associated with a frequency-dependent out-of-phase ($\chi''$) signal. This behavior indicates that **1** possibly behaves as a SMM – similar to **2**.[19] However, for the investigated frequencies $\chi''$ reaches its maximum value below 2 K, outside the experimentally accessible T-range. This is confirmed by the temperature dependence of $\chi'$, which is expected to have an inflection point when $\chi''$ displays a peak. In order to determine the peak position of $\chi''$ at different frequencies and extract the temperature dependence of the relaxation time of **1**, AC measurements below 2 K or as a function of an applied magnetic field[30] would be needed. Alternatively, higher frequencies methods like NMR[31] could be applied. Compared to **2**, complex **1** is characterized by shorter relaxation times at low temperatures thus possessing a lower blocking temperature.[19]

In summary, we have synthesized a new heterobimetallic ring MC consisting of ordered metal sites that contains a central metal ion, a first for archetypal metallacrown chemistry. In addition, **1** is only the second 12-MC-4 to consist of two different $3d$ transition metal ions. The molecule has an S = 5/2 GS and a sizeable axial ZFS on the Mn$^{\text{III}}$ ions, and the molecule possesses SMM-like behavior. Future studies will focus on synthesizing other metallacrowns with heterometallic rings and understanding how the interaction of the metal ions affects the magnetic properties.

**Supporting Information**. X-ray crystallographic information of **1** in CIF format. Experimental details, additional crystallographic details and figures, metal-ligand bond distances, BVS values, SHAPE CShM values, spin Hamiltonian and energy levels for **1**, and AC magnetic susceptibility plot (PDF). This material is available free of charge via the Internet at http://pubs.acs.org.

**Corresponding Author**

*Corresponding Authors: stefano.carretta@unipr.it; cmzaleski@ship.edu.

**Author Contributions**

The manuscript was written through contributions of all authors. All authors have given approval to the final version of the manuscript.

**Notes**

The authors declare no competing financial interest.

   **Acknowledgements.** E.G. and S.C. gratefully acknowledge financial support from PRIN Project 2015 No. HYFSRT of the Italian MIUR, from the European QuantERA 2017 project SUMO, cofounded by the Italian MUR. E.G. also acknowledges the support of the PRISM Project of the call "FIL-Quota incentivante 2019" of the University of Parma and co-sponsored by Fondazione Cariparma. C.M.Z. and A.J.L. thank the Summer Undergraduate Research Experience (SURE) program, the Faculty Professional Development Council (FPDC) Grant Program, and the Undergraduate Research Program at Shippensburg University for financial support. M.Z. thanks the National Science Foundation for financial support of the single-crystal X-ray diffractometer through the Major Research Instrumentation Program (Grant No. CHE 1625543).

SYNOPSIS

The heterometallic metallacrown (MC), $(TMA)_2\{Mn(OAc)_2[12-MC_{Mn^{III}Cu^{II}(N)shi}-4](CH_3OH)\}\cdot2.90CH_3OH$, where $TMA^+$ is tetramethylammonium, $^-OAc$ is acetate, and $shi^{3-}$ is salicylhydroximate, consists of alternating ring manganese(III) and copper(II) sites and a $Mn^{II}$ ion captured in the central cavity. DC-magnetometry characterization indicate ferromagnetic exchange between the ring metal ions, which then antiferromagnetically couple with the central $Mn^{II}$ ion. AC-susceptibility measurements reveal that metallacrown displays slow magnetization relaxation akin to single-molecule magnet behavior.

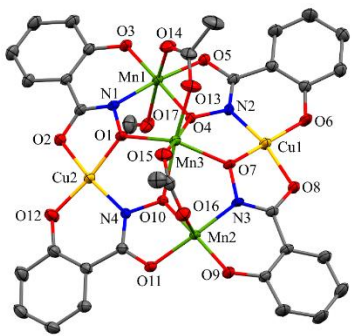